\documentclass[prd, twocolumn, nofootinbib, floatfix]{revtex4-1}

\usepackage{amsmath}
\usepackage{graphicx}
\usepackage{dcolumn}
\usepackage{bm}
\usepackage{epsfig}
\usepackage{amssymb,latexsym,mathrsfs}
\usepackage{graphicx}
\usepackage{color}
\usepackage{hyperref}

\hypersetup{
    colorlinks=true,
    linkcolor=red,
    citecolor=blue,
} 

\newcommand{\be}{\begin{equation}}
\newcommand{\ee}{\end{equation}}
\newcommand{\bs}{\begin{split}} 
\newcommand{\bea}{\begin{eqnarray}}
\newcommand{\eea}{\end{eqnarray}}

\newcommand{\ode}{\Omega_{\rm de}}

\newcommand{\gm}{G_{\rm matter}} 
\newcommand{\gl}{G_{\rm light}} 
\newcommand{\geff}{G_{\rm eff}} 
\newcommand{\geffh}{G_{\rm eff}^\Phi} 
\newcommand{\geffs}{G_{\rm eff}^\Psi} 
\newcommand{\geffsh}{G_{\rm eff}^{\Psi+\Phi}} 
\newcommand{\al}{\alpha} 
\newcommand{\kap}{\kappa}

\begin{document}

\title{Challenges in Connecting Modified Gravity Theory and Observations} 
\author{Eric V.\ Linder} 
\affiliation{Berkeley Center for Cosmological Physics \& Berkeley Lab, 
University of California, Berkeley, CA 94720, USA} 

\begin{abstract}
Cosmic acceleration may be due to modifications of cosmic gravity and to test 
this we need robust connections between theory and observations. However, 
in a model independent approach like effective field theory or a broad class 
like Horndeski gravity, several free functions of time enter the theory. 
We show that simple parametrizations of these functions are unlikely 
to be successful; in particular the approximation 
$\alpha_i(t)\propto\Omega_{\rm de}(t)$ drastically misestimates the 
observables. This holds even in simple modified gravity theories like 
$f(R)$. Indeed, oversimplified approximations to the property functions 
$\alpha_i(t)$ can even miss the signature of modified 
gravity. We also consider the question of consistency relations and the 
role of tensor (gravitational wave) perturbations. 
\end{abstract} 

\date{\today} 

\maketitle

\section{Introduction} 

The origin of cosmic acceleration is an extraordinary mystery in modern 
physics. The observation of cosmic acceleration 
\cite{perlmutter,riess,DEreview,DEreview2} 
must be connected 
to some fundamental theory beyond the current standard model of particle 
physics, but we do not know whether its origin lies in the structure of 
the quantum vacuum or an extension to Einstein's theory of gravitation. 
Considerable progress has occurred in the last decade in exploring aspects 
of modified gravity \cite{troddenreview,koyamareview,fabianreview} 
but the ability to connect theory 
and observations in a manner not highly dependent on a specific model is 
lacking in essential aspects. 

Here we examine the challenges for such a connection, and caution against 
oversimplification. Modified gravity is a much more complex arena than 
scalar field dark energy, with its one free function of time (e.g.\ the 
equation of state $w(a)$). In large part this is because of the role 
played by perturbations and the tensor sector. 

To begin, consider the case for cosmic acceleration not arising from 
modified gravity, 
e.g.\ quintessence dark energy. Here we also have challenges in connecting 
essential theory to observations, with perhaps the most information arising 
from the thawing vs.\ freezing classification of scalar fields \cite{caldlin}. 
This at least describes the steepness of the potential relative to the 
Hubble friction, and has distinct implications for whether the theory is 
approaching or departing from a cosmological constant-like (or sometimes 
de Sitter) state. Beyond that, the expansion history from dark energy, 
whether from quintessence or modified gravity, is extremely accurately 
characterized phenomenologically by two numbers \cite{depl}, $w_0$ and $w_a$, 
measures of the present and time variation of the dark energy equation of 
state. Indeed, this characterization has been shown valid to the 0.1\% 
level in the observables of distances and Hubble parameters for a wide 
range of quintessence, k-essence, modified gravity, etc.\ models. 

On the cosmic structure, i.e.\ perturbative, side of observations, 
dark energy not arising from modified gravity (or nonstandard couplings) 
has little to add: quintessence perturbations are small inside the Hubble 
scale and k-essence (noncanonical kinetic energy model) perturbations 
have little observational effect since they are suppressed by equations of 
state near $w=-1$, as observations indicate. For modified gravity effects, 
a successful, if limited, phenomenological parametrization is the 
gravitational growth index $\gamma$ \cite{lingam}, again accurately 
describing observables at the subpercent level for a variety of modified 
gravity models \cite{lincahn}. However, this has very restricted interpretable 
connection to fundamental theory. Better (pseudo)observables include 
effects not only on growth of structure, but on the deflection of light. 
These come from the nonrelativistic and relativistic modified Poisson 
equations \cite{bert11}, and can be written as effective gravitational 
coupling strengths $\gm(k,a)$ and $\gl(k,a)$, where we explicitly show 
their scale dependence (e.g.\ on the Fourier wavenumber $k$) and time 
dependence (e.g.\ on the cosmic scale factor $a$). The gravitational 
growth index $\gamma$ is directly related to $\gm$ in the scale independent 
limit. 

To connect with theory, however, we need to relate the scale and time 
dependences of these ``observables'' (we will henceforth refer to these 
quantities $\gm$, $\gl$, and their ratio, related to the gravitational 
slip function, as observables because, while not directly observable, they are 
so closely connected to observations, i.e.\ structure growth and 
gravitational lensing) to the theory -- or at least to 
phenomenological property functions $\alpha_i$ \cite{bellsaw}. 

The functional 
form of the scale dependence is a ratio of $k^0+k^2$ polynomials in many 
cases (see \cite{bert08,11084242,11091846} and the especially clear 
\cite{13021193}, but see \cite{12084163,14098284} for exceptions), 
and one simplification is that on scales below the Hubble scale 
(or more generally the sound horizon or braiding scale \cite{bellsaw}) the 
scale dependence ($k^2$ terms) is subdominant and one essentially has purely 
functions of time. 

While this seems to be considerable progress, the problem is that in order 
to know whether $\gm$ and $\gl$ have any simple parametrization of their 
time dependence one has to evaluate them from the underlying theory, 
ideally in as model independent a fashion as possible. An excellent 
framework for this is the effective field theory (EFT) of dark energy 
\cite{gubitosi,bloomfield,gleyzes,lsw}. Within EFT at 
quadratic (lowest) order there are seven free functions of time, and within 
the Horndeski class of gravity there are four free functions. The challenge 
of connecting such theory functions to realistically parametrized observables 
was highlighted in \cite{lsw}, who examined various limits. Here we go 
into greater depth and quantify the problems with oversimplification of 
the parametrization. 

In Sec.~\ref{sec:fR} we start with the workhorse of modified gravity, 
$f(R)$ theory. This corresponds to only one independent free function of 
time in the EFT formalism and so is a basic place to start in assessing 
parametrizations. We expand in Sec.~\ref{sec:early} to the four functions 
of Horndeski gravity and examine the motivation for a potential 
simplification in the early time (matter dominated) limit, deriving the 
asymptotic behavior of the property functions and observables. In 
Sec.~\ref{sec:propto} 
we identify how extending these limiting behaviors to the epoch of 
cosmological structure observations raises foundational issues, and we 
quantify the dramatic deviations that actually arise in generic 
circumstances. Section~\ref{sec:discuss} discusses the reasons why 
simplified parametrizations appear generally unviable, and solving a 
problem like modified gravity is so difficult. We conclude in 
Sec.~\ref{sec:concl} with some thoughts on further progress, while 
Appendix~\ref{sec:consist} explores the possibility of proving broad 
consistency relations that an entire class of modified gravity theories 
must obey.

\section{A One Function Case: $f(R)$ Gravity} \label{sec:fR} 

There are four free functions of time within the Horndeski class 
of gravity theories (apart from the Hubble expansion itself, $H(a)$ or 
$a(t)$, which can also be phrased in terms of an effective dark energy 
equation of state $w(a)$). It is convenient to take these functions 
to be treated in terms of property functions \cite{bellsaw}. 
The property functions describe the structure of the scalar kinetic sector of 
the theory via the kineticity $\al_K$, the tensor sector via speed of 
tensor perturbation propagation $\al_T=c_T^2-1$, the mixing of the scalar 
and tensor sectors via the braiding $\al_B$, and the running of the Planck 
mass $\al_M$. Translations between these and the EFT functions and the 
observable functions are given in, e.g., \cite{lsw}. Explicit expressions 
for $\alpha_i$ in terms of Horndeski functions are given in \cite{bellsaw}, 
and in terms of covariant Galileon $\kap$'s in, e.g., \cite{slip14}. 

In specific theories some of these functions can be zero and some can be 
redundant. In general relativity all are zero. We can start by considering 
the simplest nontrivial situation where the theory has one independent 
free function of time -- $f(R)$ is one such theory, with the only nonzero 
property function being 
$\al_M=-\al_B$ \cite{bellsaw}. Note that $\al_M$ is closely related to the 
$f(R)$ function $B(a)$, the square of the effective Compton wavelength of 
the scalaron in units of the Hubble scale. 

The property function $\alpha_M=d\ln M_\star^2/d\ln a$ describes the 
running of the Planck mass $M_\star$. Note that the strength of gravity 
is proportional to $M_\star^{-2}$ just as Newton's constant 
$G_N=M_{Pl}^{-2}$. We can write 
\bea 
M_\star^2(a)&=&M_{Pl}^2\,e^{\int_0^a d\ln a'\,\al_M(a')} 
\label{eq:mstarprop}\\ 
&=&M_{Pl}^2\,e^{\int_0^a d\ln a'\,\ode(a')\, 
[\al_M(a')/\ode(a')]}\ , 
\eea 
where the second line is in a form suggestive of an approximation where 
$\al_M(a)\propto\ode(a)$. 

We can immediately see the unfortunate consequences of such a 
proportionality approximation if it holds into the late universe. As 
dark energy dominates, $\ode\to1$, the quantity in brackets remains 
constant, but $a$ is unbounded. Thus the running Planck mass either 
goes to infinity or to zero, depending on the sign of the proportionality 
constant. Indeed we will see in the next section that if we match the 
early time behavior to evaluate the constant, then it is negative, 
forcing $M_\star^2\to0$ at late times. Thus the strength of gravity 
blows up to infinity in this approximation. 

More quantitatively, if we take a $\Lambda$CDM background expansion history, 
or a constant $w$ dark energy equation of state, we can do the integral 
analytically to find 
\be 
M_\star^2(a)=m_p^2\,\left(1+\frac{\Omega_{{\rm de},0}}{\Omega_{m,0}}\,a^{-3w}\right)^{\bar\al_M/(-3w)} \ , 
\ee 
where $\bar\al_M$ denotes the proportionality constant and subscripts 0 
indicate the present densities of dark energy and matter 
(and $w=-1$ for the cosmological constant case). As $a$ gets large, 
$M_\star^2$ is driven to zero or infinity, depending on the sign of 
$\bar\al_M$. 

The way out of this unphysical catastrophe is to break the proportionality 
\be 
\al_i^{\rm prop}=\bar\al_i\,\ode(a)\ , \label{eq:prop} 
\ee 
at some epoch. Indeed, physically we know this must happen: as the 
universe approaches a de Sitter state the running of the Planck mass 
must freeze, i.e.\ $\al_M\to0$. 

\begin{figure}[htbp!]
\includegraphics[width=\columnwidth]{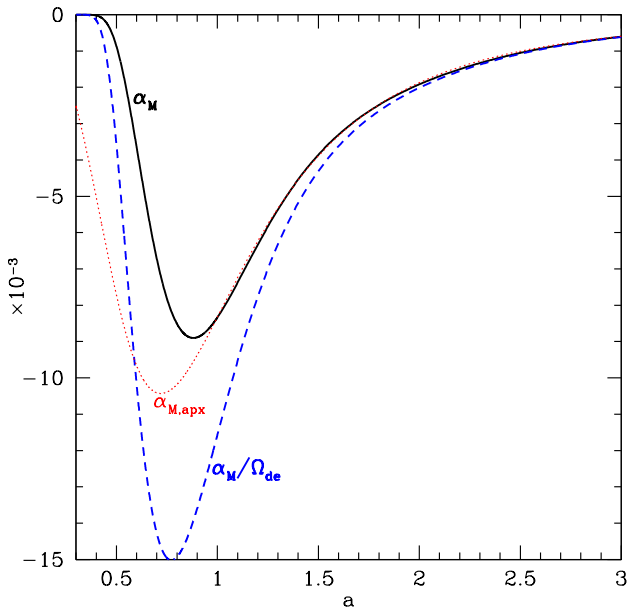} 
\caption{
The property function $\al_M$ (solid black) and its ratio relative to the 
effective dark energy density, $\al_M/\ode$ (dashed blue), is plotted vs 
scale factor $a$. 
It is well behaved and physical, with the expected early and late time limits. 
Note that the quantity $\al_M/\ode$ is definitely not well approximated by a 
constant in the recent universe (or the late time universe). 
We also show the result (dotted red) when calculating $\al_M$ using 
Eq.~(\ref{eq:mprop}). 
It gives an excellent approximation at late times but not at early times; 
the text explains why any function of $\ode$ is likely to fail during the 
observational epoch. 
} 
\label{fig:alffr} 
\end{figure}

Let us explore through the exact numerical solution of the $f(R)$ gravity 
model when the approximation that the property function (deviation from 
general relativity) is proportional to the effective dark energy, which 
we now abbreviate as {\it prop\/}, breaks down. Figure~\ref{fig:alffr} shows 
the numerical solutions for $\al_M(a)$ and $\al_M(a)/\ode(a)$, for the 
exponential $f(R)$ gravity model with $c=3$, compatible with current 
observations, given in \cite{linfr}.

We see that even for the simple $f(R)$ model that $\al_M\propto\ode(a)$ 
is a poor approximation. Indeed, $\al_M\approx10^{-10}$ at $a=0.3$ while 
$\al_M\approx10^{-2}$ at $a=1$, while $\ode$ only changes by one order of 
magnitude over this range. 
This should be no surprise: $f(R)$ gravity involves a function 
of the Ricci scalar $R$, which has a steep time dependence. Indeed, we want 
$f(R)$ to restore to general relativity rapidly in the high curvature regime. 
Due to this very steep dependence, it is difficult to see that any reasonable, 
model independent function of $\ode(a)$ will approximate $\al_M$ during 
the observable epoch $z\approx0-3$ ($a\approx0.25-1$). 

Recall that {\it prop\/}, i.e.\ that $\al_i(a)/\ode(a)=\,$constant, had the 
problematic feature that it forces $M^2_\star\to0$ at late 
times, with the consequence that the strength of gravity, $\geff$, blows 
up to infinity. Let us attempt to heal this pathology at least. 

We take as our ansatz instead 
\be 
M^2_\star=m_p^2\,\left[1+\mu\ode(a)\right] \ , \label{eq:mprop} 
\ee 
where $\mu$ is a constant. That is, instead of $\alpha_i$ deviating from 
general relativity at early times proportional to the effective dark energy 
density, instead it is the running Planck mass that has such a linear 
deviation. So at early times the Planck mass restores to general relativity, 
and at late times it freezes to a constant. The latter is what we expect 
physically in the de Sitter phase. Moreover, now $M^2_\star(a)$ is a function 
of the background expansion only at that scale factor, rather than an 
integral over all past history as in the {\it prop\/} case. 

From this we find that 
\be 
\al_M(a)\equiv\frac{d\ln M^2_\star}{d\ln a}= 
\frac{-3\mu w(a)\ode(a)\left[1-\ode(a)\right]}{1+\mu\ode(a)} \,, 
\label{eq:almstar} 
\ee 
where $w(a)$ is the effective dark energy equation of state function. 
At early times, for many modified gravity models $w(a)$ is constant so 
to first order in $\ode(a)$ we do have $\al_{M,{\rm early}}\propto\ode(a)$. 
At late times, in the de Sitter phase $\ode\to1$ and hence $\al_M\to0$, 
exactly as physically expected. Equation~(\ref{eq:almstar}) for $\al_M$, 
coming from Eq.~(\ref{eq:mprop}), is 
plotted as the dotted red curve in Fig.~\ref{fig:alffr}. 

While Eq.~(\ref{eq:mprop}) leads to a remarkably good approximation to 
$\al_M$ for times after the present, it too fails at early times. 
Indeed this rapid evolution for the gravitational coupling strength was 
discussed in terms of the ``paths of gravity'' -- the phase space diagram 
of the gravitational strength -- in \cite{linroysoc}. 

Thus, while we managed to remove a pathology and found a parametrization 
suitable for the latter half of evolution, we still do not see the way 
to a reliable parametrization for $\al_i(a)$ for observational data, even 
in this simplest case of a single free function of time in the modified 
gravity theory.

\section{Early Time Limit} \label{sec:early} 

Let us back up and understand why the simplified parametrization 
$\al_i(a)\propto\ode(a)$ seemed to be an attractive first attempt. 
We will focus on what physics can lead to such a relation, and what 
physics breaks it. 

In the early time limit we expect general relativity to be an excellent 
description of gravity, as seen from observational constraints from 
primordial nucleosynthesis and the cosmic microwave background recombination 
epoch. Not only should deviations from general relativity be small, but 
also any contributions of the effective dark energy density -- i.e.\ 
observations indicate that the universe was matter dominated (including 
radiation dominated). The impact of this on the behavior of modified 
gravity functions was discussed qualitatively by \cite{lsw} in terms of 
all the EFT functions being of the same order, as well showing how this 
arises within the specific case of covariant Galileons. Some quantitative 
behaviors for the time dependence of the observables and the property 
functions $\al_i(t)$ were also derived in \cite{gal11,slip14}. 

Within the framework of EFT, or the property functions (we now consider 
all four within the full Horndeski class as independent functions of 
time), each function is made up of an array of terms from the theory 
Lagrangian (see, e.g., \cite{jolyon}). That is, each contains terms 
depending on different numbers of times the field enters and different 
numbers of derivatives. Since each term has different dependences on the 
Hubble parameter $H/H_0$, which is large at early times, generically one 
term dominates at early times. However, as just stated, this term 
generally contributes to all the property functions, the observables, 
and the effective dark energy density. Since these are thus proportional 
to each other and the dark energy density, one has 
\be 
\al_{i,{\rm early}}\propto\ode(a) \ . \label{eq:eprop} 
\ee 

We make this relation explicit in the following, and derive the constants 
of proportionality for various Horndeski cases. However, we emphasize 
strongly that Eq.~(\ref{eq:eprop}) is only the early time limit -- 
some specific conditions for when this proportionality breaks down are 
given in \cite{lsw} and we elaborate on them here (in particular see 
Sec.~\ref{sec:propto}), as well as show when this whole ansatz is invalid 
even at early times. 

To calculate the early time relations, 
recall that for Horndeski gravity the Lagrangian consists of a sum of 
terms with the scalar field $\phi$ (and its derivatives) entering two through 
five times. The prefactors of these operators are functions $G_i$, with 
$i=2,\dots$ 5, and their derivatives, and these Horndeski functions depend 
on $\phi$ and its kinetic energy $X=\dot\phi^2/2$, i.e.\ $G_i(\phi,X)$. 
The early time behavior of $G_i$ will be determined by the leading order 
``pole'' behavior, e.g.\ the lowest power of $X$ (or $\phi$) that enters. 
Thus we will treat the early time limit in terms of 
$G_i\propto X^n$ or $\phi^m$. (In the uncoupled covariant Galileon case 
of Horndeski gravity, $G_2,\,G_3\propto X$ and 
$G_4,\,G_5\propto X^2$, with $G_4$ also having a constant part.) 

We can use the generalized Klein-Gordon equation for the scalar field 
evolution to define 
\be 
\beta\equiv\frac{2\ddot\phi}{H\dot\phi}=\frac{\dot X}{HX} \ . 
\ee 
In the early time limit, $\beta$ will go to a constant. We can evaluate 
this constant using that in general the $G_5$ term dominates at early 
times due to the number of products of the (large) Hubble parameter from 
its associated operators. Thus initially we consider $G_5\propto X^n$ (we 
consider powers of $\phi$ later). In this case 
\be 
\beta=\frac{-3(1+\dot H/H^2)(2n+1)}{2n+1+(n-1)[5+2(n-1)(n-2)]} \ . 
\label{eq:betamat} 
\ee 
Note that since terms like $n-2$ come from $G_{5XX}$, i.e.\ 
two derivatives with respect to $X$, then if $n=0$ these terms will not 
actually exist. For a background equation of state $w_b$, then 
$\dot H/H^2=-(3/2)(1+w_b)$, i.e.\ $-3/2$ for nonrelativistic matter 
domination. 

Using the known expressions for the property functions $\alpha_j$ in terms 
of $G_i$, and for $\ode$ 
in terms of $G_i$, we can solve for the early time limits of the property 
functions: 
\bea 
\frac{\al_B}{\ode}&\to& \frac{3(2n+1)}{2n+3} \longrightarrow \frac{15}{7}\\ 
\frac{\al_K}{\ode}&\to& \frac{6[7n-4+2(n-1)(n-2)]}{2n+3} \longrightarrow 
\frac{60}{7}\\ 
\frac{\al_M}{\ode}&\to& \frac{-3}{2n+3}\, 
\left[\frac{\dot H}{H^2}+\beta(n+\frac{1}{2})\right] \longrightarrow 
\frac{-9}{56} \label{eq:amearly}\\ 
\frac{\al_T}{\ode}&\to& \frac{-3}{2}\,\frac{\beta-2}{2n+3} \longrightarrow 
\frac{15}{56} \ . \label{eq:atearly} 
\eea 
Here the short arrow denotes the early time limit, and the long arrow 
denotes the further specialization to the covariant Galileon case (where 
$\beta=3/4$ in nonrelativistic matter domination). These 
constants agree with our numerical computation of the full evolution. 
Note that the first two lines do not depend on $\beta$ while the last two 
lines do; one can use Eq.~(\ref{eq:betamat}) to write those expressions 
wholly in terms of $n$. 

For the metric 
\be 
ds^2 = -\left( 1 + 2 \Psi \right) dt^2 + a^2(t) \left( 1 - 2 \Phi \right) 
d\vec{x}^2 \ , 
\ee
we can consider the effective gravitational coupling strengths appearing 
in the modified Poisson equations for non-relativistic and relativistic 
particles, e.g.\ galaxies and light, 
\bea 
\nabla^{2} \Psi &=& 4\pi a^{2} G^{\Psi}_{\rm eff} \rho_m\,\delta_ m \\  
\nabla^{2} (\Psi+\Phi) &=& 8\pi a^{2} G^{\Psi+\Phi}_{\rm eff} 
\rho_m\,\delta_m \ . 
\eea 
The quantity $\geffs$ is also called $G_{\rm matter}$ and the quantity 
$\geffsh$ is also called $G_{\rm light}$. 

The property functions can then be propagated to these and other 
``observables'' such as the gravitational slip $\eta$ and tensor wave 
speed $c_T$ \cite{slip14}; using the above early time limits, and 
specializing to the covariant Galileon limit for simplicity, 
\bea 
G^\Psi_{\rm eff,\,early}&=&1+\frac{759}{224}\ode\\ 
\eta_{\rm early}&=&1+\frac{111}{32}\ode\\ 
c^2_{T,{\rm early}}&=&1+\frac{15}{56}\ode\ . 
\eea 

Now let us consider the case where the $\phi$ dependence of $G_5$ is the 
dominant contribution. In this case one can readily find that 
\bea 
\frac{\al_B}{\ode}&\to& \frac{4}{3}\\ 
\frac{\al_K}{\ode}&\to& 2\\ 
\frac{\al_M}{\ode}&\to& -\frac{1}{3} 
\left[\beta+(m-1)\frac{\dot\phi}{H\phi}\right] 
\longrightarrow -\frac{\beta}{3}\\ 
\frac{\al_T}{\ode}&\to& \frac{2}{3} \ . 
\eea 
Here the long arrow denotes specialization to the derivatively coupled 
covariant Galileon, where $G_5\sim c_G\phi$. However in this case $\beta$ 
is no longer given by Eq.~(\ref{eq:betamat}). Instead, $\beta=6w_b$. Indeed, 
from Eq.~50 of \cite{gal11} we find that in the nonrelativistic matter 
early time limit with the $c_G$ term dominating, 
$X=\,$constant and hence $\beta=0$, so $\alpha_M=0$. 

One can carry out the same analysis if another term than the expected $G_5$ 
dominates the Horndeski Lagrangian at early times. 

The basic rule is that as long as the same term 
dominates for both $\al_i$ and $\ode$, one will obtain their proportionality 
in the early time limit. The next section will go beyond the early time 
limit, but first we should look for any exceptions to the early time 
proportionality arising from a mismatch between the terms entering 
$\ode$ and each $\al_i$. We find that indeed $\al_B$, $\al_K$, 
and $\al_T$ all lack certain terms that $\ode$ has, while $\al_M$ has a 
term that $\ode$ lacks. 

For example, $\al_B$ is lacking the term $G_{3\phi}$ that $\ode$ has, so 
if the theory is arranged (possibly fine tuned) to make this dominant at 
early times, then $\al_B/\ode\to0$. A similar situation occurs for 
$\al_K$ and $\al_T$ when $G_{4\phi}$ is dominant (and for $\al_T$ when 
any $G_3$ term dominates). These results will have important implications 
in the next section. 

For $\al_M$, there is an extra term involving $\dot\phi^3 G_{5\phi\phi}$. 
For a leading order behavior of $G_5\sim\phi^m$, this term involves 
$m(m-1)(\dot\phi^3/\phi^2)G_5$ and so one can have 
\be 
\frac{\al_M}{\ode}\sim \frac{\dot\phi}{H\phi}\quad {\rm or}\quad 
\left(\frac{\dot\phi}{H\phi}\right)^2 \ , 
\ee 
whichever is dominant, unless $m=0,\,1$. The ratio therefore will in general 
either go to zero, if $\dot\phi/(H\phi)$ is small, 
giving a similar problem as with the other $\al_i$, or diverge, 
if $\dot\phi/(H\phi)$ is large, giving a new problem. 

Thus, {\it prop\/} proportionality is not even guaranteed at early times, 
while we found in Sec.~\ref{sec:fR} that it fails during the observational 
epoch in the recent universe for the well known, simple case of $f(R)$ 
gravity. We investigate further in the next section.

\section{Parametrizing Property Functions} \label{sec:propto}

\subsection{Limits to linear proportionality} \label{sec:linlim} 

The early time limit is the only case where the behavior of the property 
functions, and observables, can be calculated analytically. As seen in 
the previous section, this showed that at early times $\al_i$ was almost 
always proportional to $\ode$. However, there are three important caveats: 

\begin{itemize} 
\item{We saw in Sec.~\ref{sec:early} that for some theories the constant 
of proportionality was either zero or infinity. These give behavior at 
later times that is clearly invalid (or trivial) within the proportionality 
approximation.}  
\item{In the de Sitter limit one must have $\al_M=0$ so proportionality 
must break down for this function.} 
\item{The behavior actually deviates from the early time asymptote at 
quite early times.} 
\end{itemize} 

The discussion in \cite{lsw} makes clear that linear proportionality 
breaks down not when $\ode$ becomes appreciable compared to unity 
(i.e.\ at redshift $z\lesssim1$) but when 
$H/H_0$ is no longer much greater than one, i.e.\ at redshift $z\approx10$ 
(for a $\Lambda$CDM background, $H/H_0=20$ at $z=10$). Remember, the 
physics comes from the interrelation of the multiple terms in the Lagrangian 
with different powers of $H$. Furthermore, we will 
see that even at $z=10$, the observable functions $\gm$ and $\gl$ 
calculated from the combinations of the property functions are poorly 
approximated by using a linear $\al_i\propto\ode$ relation. 
Note though the very useful modified gravity Boltzmann code 
{\tt hi\_class} \cite{hiclass} provides Eq.~(\ref{eq:prop}) as a default 
parametrization.

\subsection{Tracker trajectories} \label{sec:track} 

Can we force the linear proportionality to hold longer, despite the 
strong physical basis for the breakdown? If we can freeze the relation 
between the Lagrangian terms, preventing their natural evolution relative 
to each other, then this may be possible (although such a construction 
certainly breaks model independence and raises the specter of fine tuning). 
While the terms differ in powers of $H$, they also differ in powers of 
other dynamical variables, and so one could impose conditions on their 
combination to force the terms in lockstep. 

This is what ``tracker'' models do: they fix $H^2X=\,$constant for all 
times. When this holds, then all the Horndeski $G_i$ are the same order, 
despite their differing powers of $H/H_0$. This certainly gives up 
model independence by narrowing to a specific subclass\footnote{Note 
that \cite{barreira} claims that all other 
regions of covariant Galileons are observationally unviable but their analysis 
only concerns cubic Galileons, which have only a single term other than the 
kinetic one and so lack the freedom of the full covariant Galileon, let 
alone the Horndeski class.}. 
Moreover this then implies that 
$H^2\rho_{\rm de}=\,$constant \cite{deftsu10}. That is a dramatic 
imposition. 

\begin{figure}[htbp!]
\includegraphics[width=\columnwidth]{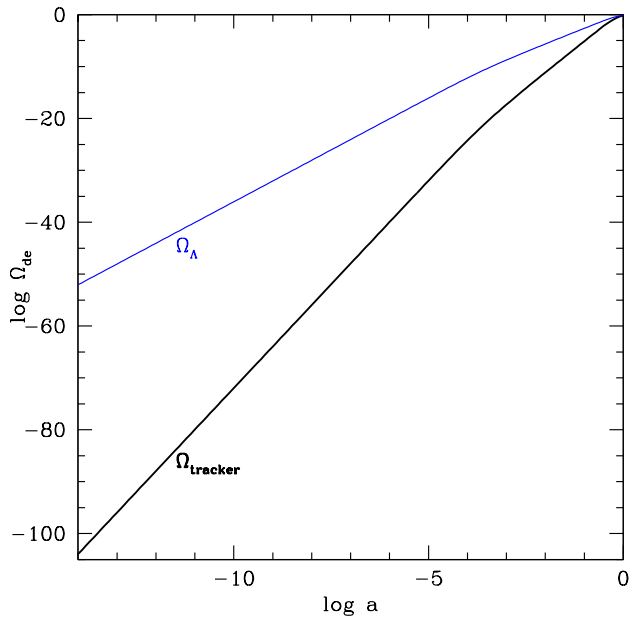} 
\caption{The fractional energy density in dark energy is plotted vs the 
log of the scale factor from $a=10^{14}$ to the present, for the tracker 
assumption (thick, black curve) and the cosmological constant (thin, blue 
curve). Assuming that 
the behavior follows the tracker at early times (rather than only at 
late times as in the generic physics scenario) imposes severe fine tuning, 
well beyond the fine tuning of the cosmological constant problem.} 
\label{fig:finetune} 
\end{figure}

Recall that 
\be 
\ode(a)\propto \frac{\rho_{\rm de}}{H^2} \ , 
\ee 
so the tracker condition forces 
\be 
\ode(a)\propto \left(\frac{H(a)}{H_0}\right)^{-4} \ . \label{eq:finetune} 
\ee 
By contrast, 
\be 
\Omega_\Lambda(a)\propto \frac{\rho_\Lambda}{H^2} \propto 
\left(\frac{H(a)}{H_0}\right)^{-2} \ . 
\ee 
Figure~\ref{fig:finetune} illustrates the implications of this. 
The tracker model is more fine tuned than even a cosmological constant,  
by $(H_0/H)^2$. We see that if one extended this behavior back to the 
Planck scale, then the 
tracker model has a fine tuning of $10^{-240}$ in contrast to the 
cosmological constant's $10^{-120}$. 
Some articles in the literature, e.g.\ \cite{renk,barreira}, use this 
condition back to $z=10^{14}$, where Fig.~\ref{fig:finetune} shows the 
fine tuning is at the level of $10^{-104}$, or $10^{52}$ times more severe 
than the cosmological constant at that redshift. These results agree 
completely with Figure~11 of \cite{barreira}, which only plots the 
density back to $a=10^{-3}$.

The physics behind the approach to the $H^2 X=\,$constant behavior 
is the same as that causing cosmic acceleration. One can immediately 
recognize that when this is written in the form 
$H^2\rho_{\rm de}=\,$constant this is merely the de Sitter attractor, 
when $H$ and $\rho_{\rm de}$ become constant. The natural epoch for the 
approach to the tracker behavior of $H^2X=\,$constant is simply 
$z\approx0$; the general physics says that, barring fine tuning, 
one would expect the behavior not to hold before the present. 
This is borne out by numerical solution of the evolution equations. 

Suppose one did allow the severe fine tuning required by 
Fig.~\ref{fig:finetune}. This not only gives up model independence by 
narrowing to a highly specific subregion of theory space, but also has 
theoretical issues: the condition $H^2X=\,$constant (which motivates 
Eq.~\ref{eq:prop}) forces the tensor perturbation propagation speed 
(for the uncoupled or derivatively coupled Galileon) to be less than 
the speed of light, $c_T^2<1$, leading to a gravi-Cherenkov catastrophe 
\cite{caves,nelson,stoica,kimura}. 

Thus the means of preventing the early breakdown of linear proportionality 
by forcing $H^2 X=\,$constant does not appear to be a generally viable 
solution.

\subsection{Numerical solutions of evolution} \label{sec:numevo} 

Let us examine the exact numerical solutions of the property functions and 
observable 
functions to investigate the question of reasonable parametrizations. 
The relations between the property and observable functions are 
\cite{bellsaw} 
\be 
\frac{\geffh}{G_N}= \frac{2m_p^2}{M_\star^2} 
\frac{[\alpha_B(1+\alpha_T)+2(\alpha_M-\alpha_T)]+\alpha_B'}{(2-\alpha_B)[\alpha_B(1+\alpha_T)+2(\alpha_M-\alpha_T)]+2\alpha_B'} \ , \label{eq:geff}
\ee 
where prime denotes $d/d\ln a$. 
The gravitational slip $\bar\eta=\gm/\gl=\geffs/\geffsh$ (note that 
$\eta=\geffs/\geffh=\bar\eta/(2-\bar\eta)$) is given by 
\be
\bar\eta=
\frac{(2+2\alpha_M)[\alpha_B(1+\alpha_T)+2(\alpha_M-\alpha_T)]+(2+2\alpha_T)\alpha_B'}{(2+\alpha_M)[\alpha_B(1+\alpha_T)+2(\alpha_M-\alpha_T)]+(2+\alpha_T)\alpha_B'} \ , \label{eq:etafull} 
\ee 
and the tensor wave speed is 
\be 
c_T^2=1+\alpha_T \ . 
\ee 

\begin{figure}[htbp!]
\includegraphics[width=\columnwidth]{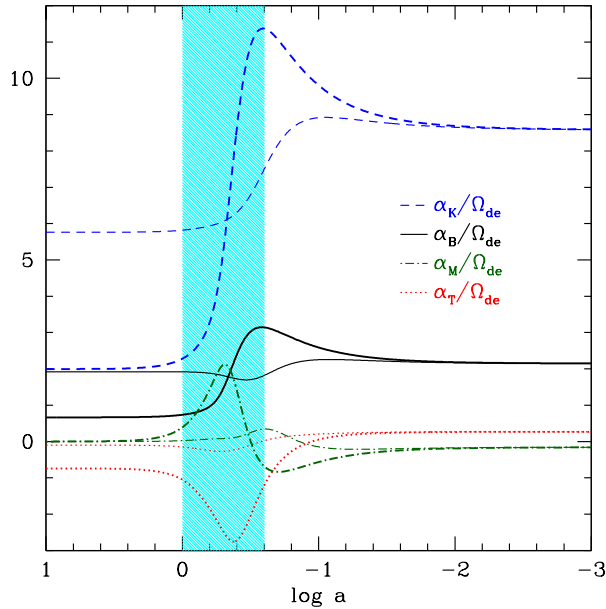} 
\caption{
The time dependence of the property functions divided by the effective 
dark energy, $\alpha_i(a)/\ode(a)$, are exhibited 
for exact solutions corresponding to case 1 (the thicker curves, with larger 
variation) and case 2 (the thinner curves, with smaller variation). 
They are not constant, as the {\it prop\/} approximation of 
Eq.~(\ref{eq:prop}) assumes. Such an approximation is particularly inaccurate 
during the key observational epoch $z=0-3$ ($\log a\approx -0.6-0$, shaded). 
} 
\label{fig:aode} 
\end{figure}

For definiteness in the numerical exploration, we here work with uncoupled 
covariant Galileon gravity; recall we showed the results for $f(R)$ gravity 
in Sec.~\ref{sec:fR}. We calculate for the models of Fig.~6 and 
Fig.~4 of \cite{gal11} (slightly adjusting $c_2$ to obtain 
$\Omega_{{\rm de},0}=0.713$), which exhibit very different de Sitter limits 
for the observable functions. We call these case 1 and case 2. 
Figure~\ref{fig:aode} plots the property 
functions $\al_i(a)$, divided by $\ode(a)$ to examine whether such a 
ratio is really constant for all times. The vertical shaded region 
highlights the region $z=0-3$ where cosmic structure data exists, and 
we are particularly interested in an accurate parametrization (taking the 
constant of proportionality to be a fit parameter, rather than fixed to 
its analytic, early time value). The quantities $\al_i/\ode$ are certainly 
seen to be not well approximated as constant (regardless of the value of 
the constant), especially over this range.

Trying to define a constant of proportionality by taking the early time 
(high redshift) value -- where proportionality does hold -- can even give 
the wrong sign during the observational epoch: see the $\al_M$ and $\al_T$ 
curves. Note that the physics of 
Horndeski gravity requires $\al_M=0$ in the de Sitter limit, while the 
linear proportionality approximation of Eq.~(\ref{eq:prop}) violates this 
for any nonzero constant. Finally, 
defining the constant of proportionality by an average over cosmic history 
(hence not obeying either the early time or late time limits) could greatly 
reduce the sensitivity to observing deviations from general relativity, as we 
see next. 

\begin{figure}[htbp!]
\includegraphics[width=\columnwidth]{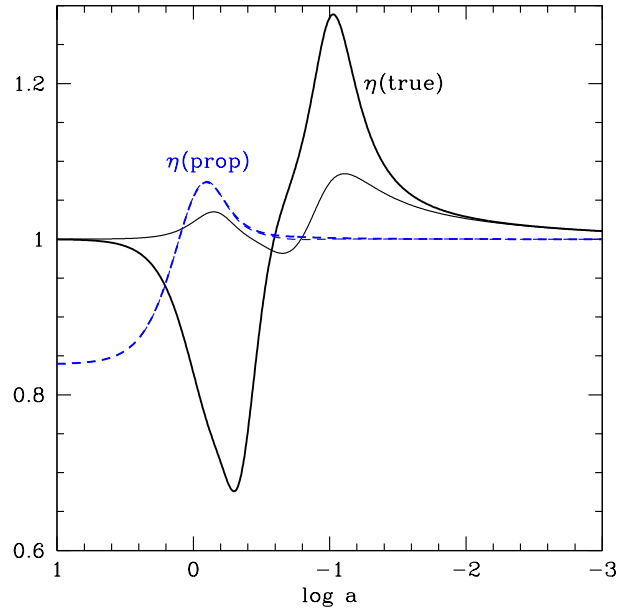} 
\caption{
The time dependence of the gravitational slip $\eta$ is compared for the 
exact solution (solid curves) and the $\alpha_i\propto\ode(t)$ approximation 
(dashed curves). The {\it prop\/} approximation gives completely different 
and inaccurate results for the physics. While the true behavior depends on 
the differing parameters of the theory (shown for the same two cases as 
Fig.~\ref{fig:aode}), the {\it prop\/} approximation has nearly identical 
behavior since the models have nearly the same expansion history. The 
{\it prop\/} approximation also underestimates the deviation from general 
relativity, possibly missing detection of modified gravity, and has an 
incorrect de Sitter asymptote. 
} 
\label{fig:etaapx} 
\end{figure}

In Fig.~\ref{fig:etaapx} we calculate the gravitational slip observable 
function $\eta$. In addition to the numerical solution we show the 
predictions for the same models using the linear proportionality 
approximation. Note that because the 
background expansion histories for cases 1 and 2 are very similar, the 
{\it prop\/} approximation, which is a function only of the background, 
delivers nearly the same observable function for each. However the true 
solutions 
show highly differing behaviors for the two cases. Moreover, at $z\gtrsim1$, 
{\it prop\/} shows almost no deviation from general relativity (below 
1\% for $z>2$, below 3\% for $z=1-2$), while the true solutions have 
considerably larger deviations. Thus, using the linear proportionality 
assumption can miss even quite dramatic signatures 
of modified gravity. Finally, as we saw for $\alpha_M$, the 
{\it prop\/} approximation does not give the physically required de Sitter 
property that $\eta=1$ for Horndeski gravity.

\begin{figure*}[htbp!]
\includegraphics[width=\columnwidth]{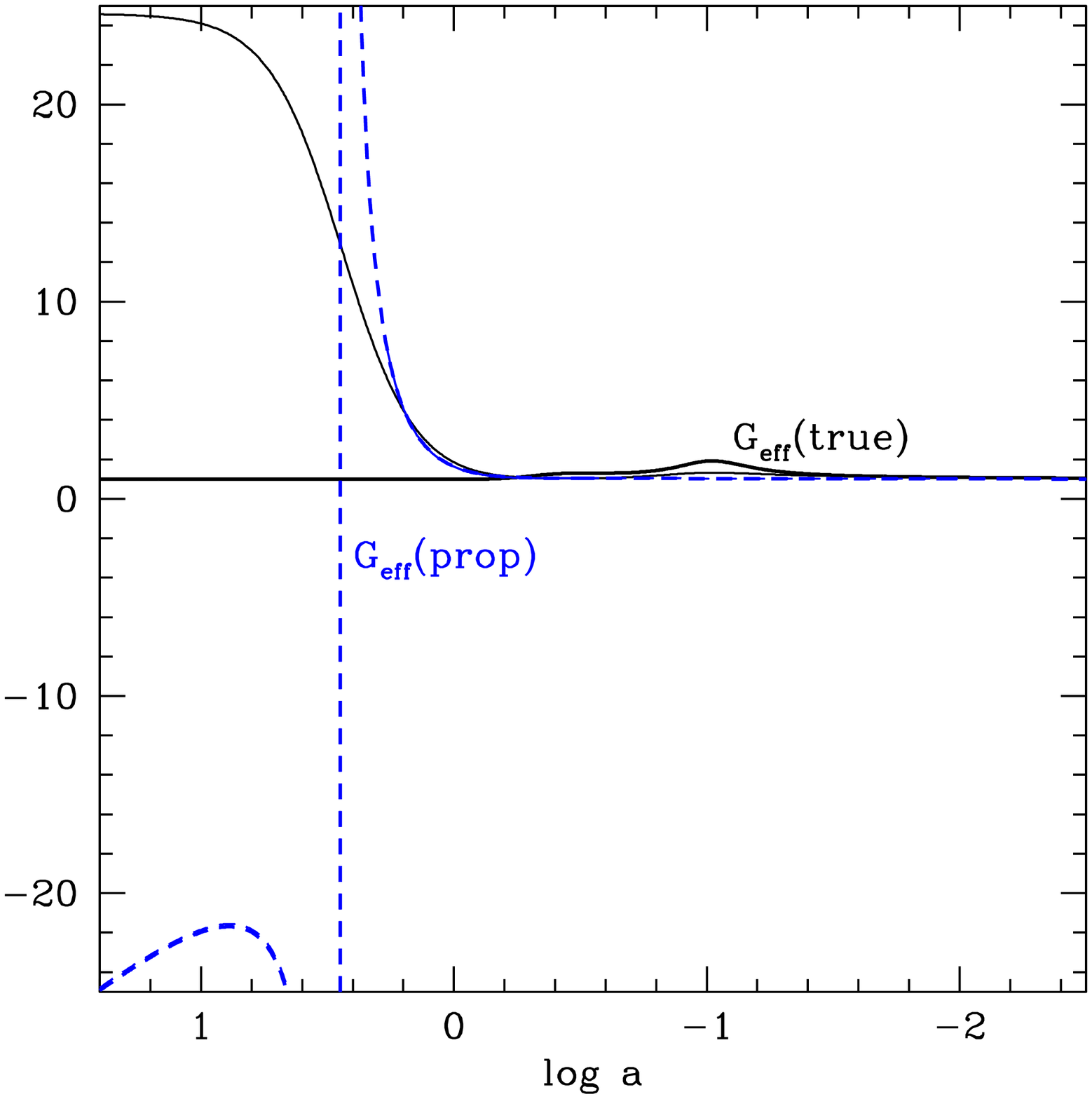} 
\includegraphics[width=\columnwidth]{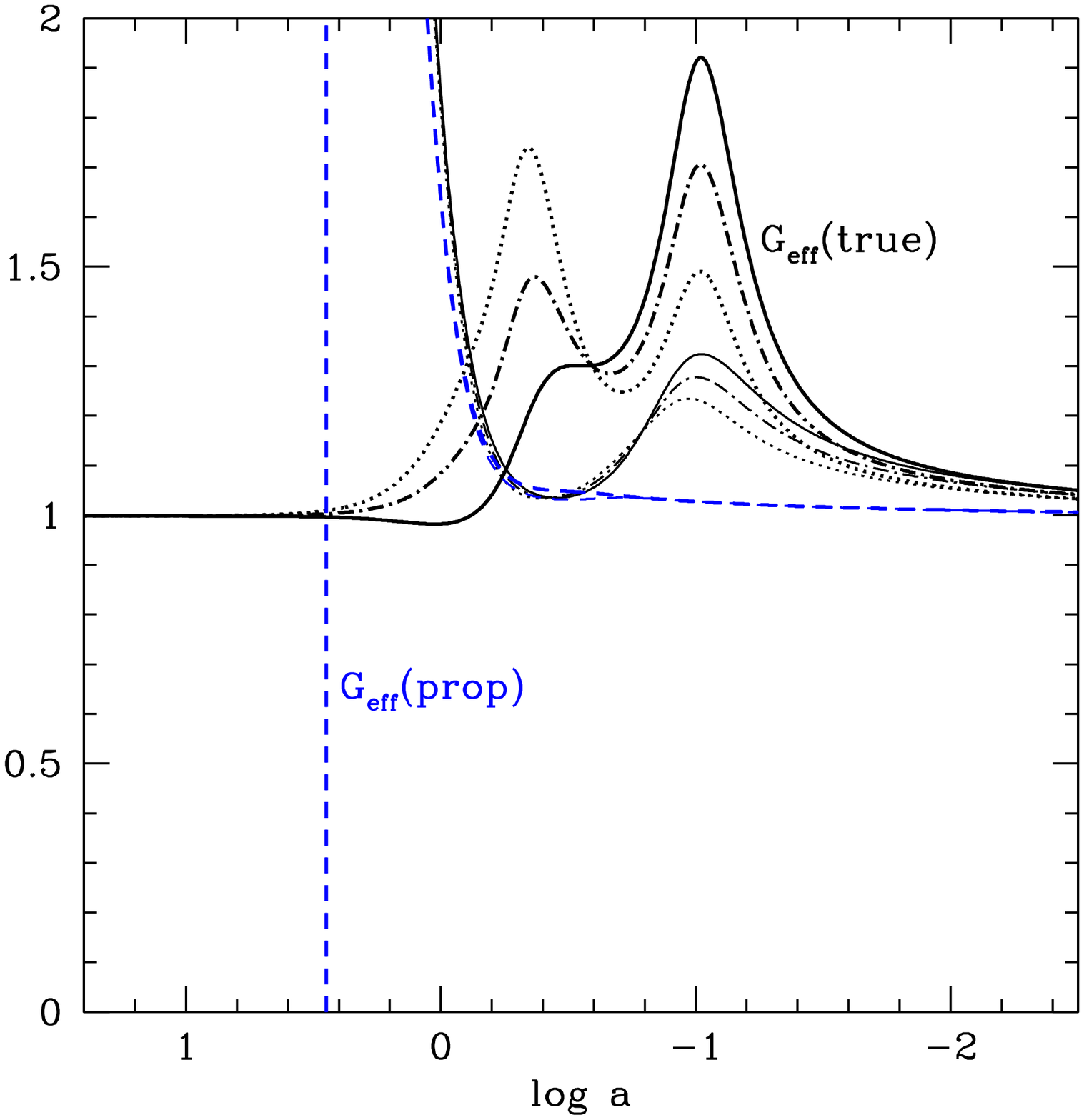} 
\caption{
The time dependence of the gravitational coupling $G_{\rm eff}$ is compared 
for the exact solution (solid curves) and the $\alpha_i\propto\ode(t)$ 
approximation (dashed curves). 
The left panel shows the global behavior of $\geffh$ while the right panel 
zooms in on the detail around the observational epoch and also shows 
$\gm$ (dotted curves) and $\gl$ (dot-dashed curves). 
The linear proportionality approximation gives completely different 
and inaccurate results for the physics. While the true behavior depends on 
the parameters of the theory (shown for the same two cases as 
Fig.~\ref{fig:aode}), the {\it prop\/} approximation has nearly identical 
behavior since the models have nearly the same expansion history. The 
{\it prop\/} approximation also underestimates the deviation from general 
relativity during the observable epoch, possibly missing detection of 
modified gravity, and has an incorrect, and divergent, de Sitter asymptote. 
} 
\label{fig:geffapx} 
\end{figure*}

Next we consider the gravitational coupling $\geff$. The left panel of 
Fig.~\ref{fig:geffapx} shows the true, numerical solutions and {\it prop\/} 
predictions for the two cases. The right panel zooms in on the detail within 
the observational epoch. Again we see that {\it prop\/} almost entirely 
misses the modified gravity signal, cannot distinguish between the two 
different cases, and has a pathological late time limit where $\geff\to 
-\infty$.

\section{Fitting Modified Gravity} \label{sec:discuss} 

If linear proportionality as a method for parametrizing the time 
dependence of the property functions is not generically valid, is 
there some other low dimensional (few parameter) approximation? 

Figures~\ref{fig:aode}, \ref{fig:etaapx}, \ref{fig:geffapx} exhibit the 
challenge of parametrizing modified gravity with a simple time dependence 
for either the property functions or observational functions. Even at 
early times when $\al_i/\ode$ does not appear to be far from constant, 
the small deviations have a large impact on the observables. For example, 
at $a=0.1$ ($z=9$) the property functions $\al_i/\ode$ deviate from 
{\it prop\/} by 5\%, 4\%, 9\%, and 24\% for subscripts B, K, M, T. 
Deviations from general relativity in the observable functions $\eta$ 
and $\geffh$ of 7\% and 32\% respectively -- just for case 2, the smaller 
variation case -- are missed by the constant proportionality approximation. 
Recall that $\eta=1$ to within 1\% for $z>2$ according to the {\it prop\/} 
approximation. 

Note that any attempt to make the property functions $\al_i(a)$ follow 
the effective dark energy density -- whether through linear proportionality 
or a more complicated function -- has a disadvantage from a physics 
perspective. One hallmark of modified gravity is that growth does not 
follow expansion, so attempting to make the growth purely a function of 
expansion does not seem to follow this. The four (or more) free functions 
of time within EFT are in addition to $H(a)$, or $\ode(a)$, and it 
restricts highly the physics if they are all forced to be strictly 
dependent on it. 

Within the observational epoch the behavior tends to be quite complicated. 
Also, note that in general we need to know values of the property functions 
or observable functions at all times before the present. 
The quantity $M_\star^2$ that enters the gravitational strength $\geff$, 
and hence the growth, requires an integral 
over all past history (hence the deviations discussed above at $z\ge2$ 
are important for lower redshift observations as well). 
Thus even a three parameter parametrization for each $\al_i$ such as 
using binned values for $z\in[0,1]$ and $[1,3]$ and constant proportionality 
at earlier times does not work well we find. 
Approximate functional forms, bins, or principal components all fall short 
because of the complexity of the relation between the theory and 
observables; as stated in \cite{lsw}, these relations are at best the 
ratios of sums of products of ratios of sums of functions. 

The failure of parametrizations should not be a huge surprise. The degrees 
of freedom in a general model are too manifold. 
For example, although all Horndeski models with a de Sitter late time 
behavior have the same background expansion and $\eta=1$ there, the values of 
$\geff$ can widely vary, as seen in Fig.~\ref{fig:geffapx}. Similarly, 
while any Horndeski models with the same dominant function, e.g.\ 
$G_5(\phi,X)$, and functional form at early times will have the same 
values of $\al_i/\ode$ then, at observable times the interplay between all 
the terms is important and cannot be made model independent. 
Modified gravity cannot be forced into a few simple numbers without 
restricting to a specific model or perhaps the benefit of some new 
theoretical insight. 

Even for the simplest case of one function of time, as seen in 
Fig.~\ref{fig:alffr} for $f(R)$ gravity, the form of the numerical solutions 
give no expectation that a simple low order polynomial can capture the 
richness of the theory, let alone be model independent. We emphasize 
that this case was wholly observationally viable, so the complicated 
time dependence is not a matter of a bizarre area of model space, but 
rather is generic.

\section{Conclusions} \label{sec:concl} 

Modified gravity leading to cosmic acceleration is a much richer field 
than envisioned even a few years ago. The early models like DGP 
gravity with a single number (the crossover scale) or $f(R)$ gravity with 
a time dependent scalaron mass as described by a single power law index of 
scale factor have much less freedom compared to even the 
quite restricted covariant Galileon theories with constant coefficients, 
let alone the Horndeski class or EFT with their several free functions 
of time. 

This complexity, in both the theory and its connections to observables, 
means that accurate 
approximations to the observables -- being ``ratios of sums of products of 
ratios of sums of functions'' -- are rare. We derive analytic limits in 
the early time, matter dominated regime for general classes of Horndeski 
gravity, and show under what conditions they appear. 

These early time approximations, however, break down dramatically even at 
redshifts $z\approx10$, let alone in the heart of the observable epoch. 
Even percent level deviations in the property functions $\al_i(a)$ 
can lead to large misestimations in observable properties. In particular, 
we demonstrate that taking them 
proportional to the effective dark energy density, $\al_i(a)/\ode(a)\propto$ 
constant can lead to unphysical behavior and fine tuning and can 
miss significant signatures of departure from general relativity. This 
last property is perhaps most damaging: misestimation could just give a 
false alert, but lack of an alert will miss essential physics \cite{xkcd}. 

To meet the challenge of connecting theory and observations, we need 
some parametrization that can prove itself accurate on at least broad 
swathes of theories in the literature. The numerical solutions we have 
shown for $f(R)$ and covariant Galileon gravity, demonstrating the 
complexity of the evolution, indicate this may be a difficult task. In 
a real sense this is no surprise: the hallmark of modified gravity is 
that the physics of growth does not simply follow the expansion history, 
e.g.\ $\ode(a)$. 

If a nearer term goal is merely an alert that general relativity may not 
be matching observations, then bins in scale and time of $\gm$ and 
$\gl$, proposed in \cite{10021962,10080397}, work well. Moreover, they 
would give some indication of how the breakdown occurs, i.e.\ the trend 
in space and time variation. While the lack of an elegant parametrization 
such as exists for the background expansion (e.g.\ dark energy equation of 
state) or even simple linear growth (e.g.\ the gravitational growth index) 
is disappointing, it also points up the richness of the problem of modified 
gravity. 
In Appendix~\ref{sec:consist} we comment on a conjecture for a general 
consistency relation between observables that could apply to 
wide classes of modified gravity theories. 

At the same time, we should seek 
new gravitation theories that are neither overly simplified and so lacking 
model independence nor complicated but observationally unviable.

\acknowledgments 

I thank Tessa Baker, Alexandre Barreira, Jerome Gleyzes, and Miguel 
Zumalac{\'a}rregui for helpful discussions. Some of this work was performed 
at the Aspen Center for Physics, which is supported by NSF grant PHY-1066293. 
This work is supported in 
part by the U.S.\ Department of Energy, Office of Science, Office of High 
Energy Physics, under Award DE-SC-0007867 and contract no.\ DE-AC02-05CH11231.

\appendix

\section{Consistency Relations} \label{sec:consist} 

As an alternative method to that taken in the main text, a broader 
though less detailed approach is to consider the observables 
without any parametrization. Are there any properties of them, or 
relations between them, for which an observational constraint could rule 
out an entire class of theories? One interesting conjecture, recently 
put forth by \cite{pogosil}, was that 
the deviation from general relativity of one of the observables (i.e.\ 
$\gm$ or $\gl$, which they call $\mu$ and $\Sigma$), either positive 
or negative at some instant of time, could not have a deviation of opposite 
sign in the other quantity. The intriguing concept is that an observational 
violation of such a consistency relation would effectively rule out all 
Horndeski models. Unfortunately no proof is given but rather an assertion 
of likeliness. Let us briefly examine the expressions for the 
deviations and see if such unlikelihood is obvious. 

To make the expressions as simple as possible, consider a subclass of 
Horndeski theory called covariant Galileons. If the consistency relation 
is not obvious for the simple case, then any obviousness for the general 
Horndeski class should be more difficult to see. The expressions for 
$\gm$ and 
$\gl$ are given in \cite{gal11} (there called $\geffs$ and $\geffsh$). 
If the deviations from general relativity $\gm-1$ and $\gl-1$ have the 
same sign (note that all gravitational couplings are here normalized by the 
strength in general relativity, i.e.\ Newton's constant), then their 
ratio must be positive: 
\be 
R\equiv\frac{\gl-1}{\gm-1}>0 \ . \label{eq:consist} 
\ee 

Writing this out for covariant Galileons, 
\begin{widetext} 
\be 
R=\frac{\kap_6(2\kap_3+\kap_4)-\kap_1(2\kap_1+\kap_5)-\kap_5(\kap_4 \kap_1 
-\kap_5 \kap_3)+\kap_4(\kap_4 \kap_6-\kap_5 \kap_1)}{4(\kap_3 \kap_6-\kap_1^2) 
-\kap_5(\kap_4 \kap_1-\kap_5\kap_3)+\kap_4(\kap_4 \kap_6-\kap_5 \kap_1)} \ , 
\ee 
\end{widetext} 
where $\kappa_i$ in turn are given by sums of many terms (see \cite{gal11}). 
It does not appear obvious that $R>0$ is required, or even highly likely
(and of course the relation in the full Horndeski class is even more 
complicated). 

While a specific model with $R<0$ is not easily identified, consider instead 
\be 
R_\Phi\equiv\frac{\gl-1}{\geffh-1}>0 \ , 
\ee 
i.e.\ where $\geffh$ rather than $\geffs$ is used in the denominator. 
A violation of this relation is shown in Sec.~\ref{sec:propto}, and the 
expression for $R_\Phi$ is no more complex or substantially different from 
that for $R$. 

The consistency relation in terms of $G_{\rm matter}$--$G_{\rm light}$ 
may indeed hold, but nothing in the equations 
obviously seems to require this. A firm proof of a consistency condition 
such as conjectured in \cite{pogosil} would be highly interesting.


\end{document}